\PassOptionsToPackage{unicode}{hyperref}
\PassOptionsToPackage{hyphens}{url}
\documentclass[
]{article}

\usepackage{amsmath,amssymb}

\usepackage{tikz}
\usetikzlibrary{fit, positioning, shapes.geometric, decorations.pathreplacing, calc}
\usepackage{tikz-cd}

\tikzset{stretch/.initial=1}
\newcommand\drawloop[4][]%
{\draw[shorten <=0pt, shorten >=0pt,#1]
($(#2)!\pgfkeysvalueof{/tikz/stretch}!(#2.#3)$)
let \p1=($(#2.center)!\pgfkeysvalueof{/tikz/stretch}!(#2.north)-(#2)$),
  \n1= {veclen(\x1,\y1)*sin(0.5*(#4-#3))/sin(0.5*(180-#4+#3))}
  in arc [start angle={#3-90}, end angle={#4+90}, radius=\n1]%
}

\usepackage[margin=1.5in]{geometry}

\makeatletter
\@ifundefined{KOMAClassName}{% if non-KOMA class
  \IfFileExists{parskip.sty}{%
    \usepackage{parskip}
  }{% else
    \setlength{\parindent}{0pt}
    \setlength{\parskip}{6pt plus 2pt minus 1pt}}
}{% if KOMA class
  \KOMAoptions{parskip=half}}
\makeatother

\usepackage{longtable,booktabs,array}
\usepackage{calc} % for calculating minipage widths
\usepackage{subcaption}
\usepackage{graphicx}
\makeatletter
\def\fps@figure{htbp}
\makeatother
\usepackage{adjustbox}

\makeatother

\usepackage{hyperref}
\hypersetup{
  pdftitle={Grounded Computation \& Consciousness},
  pdfauthor={Ryan Williams},
  hidelinks}

\usepackage{url}
\usepackage{natbib}
\usepackage{mathtools}
\DeclareMathOperator*{\argmin}{arg\,min}
\DeclareMathOperator*{\argmax}{arg\,max}

\title{Grounded Computation \& Consciousness}
\makeatletter
\providecommand{\subtitle}[1]{% add subtitle to \maketitle
  \apptocmd{\@title}{\par {\large #1 \par}}{}{}
}
\makeatother
\subtitle{A Framework for Exploring Consciousness in Machines \& Other Organisms}
\author{ Ryan
Williams\vspace{0.05in} \\ \newline\normalsize\url{ryan@cognitivemechanics.org} }
\date{September 2024}

\begin{document}
\maketitle
\begin{abstract}
  Computational modeling is a critical tool for understanding consciousness, but is it enough on its own?
  This paper discusses the necessity for an ontological basis of consciousness, and introduces a formal framework
  for grounding computational descriptions into an ontological substrate.
  Utilizing this technique, a method is demonstrated for estimating the difference in qualitative experience between
  two systems.
  This framework has wide applicability to computational theories of consciousness.
\end{abstract}

Computation is a critical instrument in the scientific exploration of the mind.
The successes it has attained in modeling functions of the human intellect are unrivaled by any other method, and there
are no apparent candidates.
Any functional property of a system's structure or behavior should be amenable to being approximated in terms of such
systems in principle.
This recognition of its unique power also demands that we understand which questions it can help us answer
and where its limitations lie.
Among a host of other puzzles that the science of the mind raises, we should aim to give principled and evidenced
accounts for whether foreign systems have meaningful conscious experiences.

Functionalist and computationalist approaches to consciousness have a long history of traction from perspectives in a
variety of fields.
These range from speculative statements that present models may already be conscious (or that they would become
conscious with the appropriate changes in architecture) to explicit algorithmic (or pseudo-algorithmic) proposals
which purport to account for conscious experience.
The appeal of the functional perspective is clear: if computation is all that's needed, we already have the tools
required.

Notwithstanding this obvious attraction, I think we can give simple, well-motivated
arguments that computation requires an ontological basis to give a full account of consciousness.
The concrete feelings of our subjective experience demand a concrete substrate.
They are the furthest things from the abstract we have known to us.
There is a gap between using computational methods to model or identify conscious phenomena, and the
assertion that consciousness is \emph{itself} a computation.
I believe the evidence available indicates that the ontology—the underlying substrate—of a potentially conscious system
plays a vital role that can't be discarded.\footnote{The specific ontology can be interpreted according to a number of metaphysical positions.}

Far from being a critique of computational methods, the thesis of this paper is that we need to acknowledge the
limitations of abstract descriptions in order to harness them fully.
The paper begins with an exploration of the principles that compel us to draw from outside of computation to explain
conscious phenomena.
With an understanding of those limitations in place, we begin reassembling an approach with new tools
(including computational ones) that can help extend the reach of computational methods right to the brink of their
utility.

We can use computation itself to show us as much about the things it \emph{can't} explain as the things that it can.

Many of the general ideas presented in this paper are not new.\footnote{Among other publications cited at relevant
points in the paper, \cite{searle1990} and \cite{maudlin1989} each explore a number of compelling arguments.}
My own previous work\footnote{\cite{williams2024} and \cite{williams2022}} gives formal descriptions of certain
structural and functional properties found within conscious experience with explicit mappings to a substrate.
But I believe the new formal techniques introduced for grounding computational systems within a specific ontology,
as well as the suggested method for estimating the difference in qualitative experience between two systems, are
interesting directions to follow and incorporate with other work.

\subsection{Computational Accounts of Consciousness}\label{subsec:sampling}

The common feature of all computational accounts of consciousness is what I call \emph{The Computational Thesis}:\label{computational-thesis}

\begin{quote}
  \textbf{The Computational Thesis:} Any system that can be interpreted as being described by algorithm $X$ is conscious.
\end{quote}

The specific algorithm differs from theory to theory, with some overlap, and is in most cases only loosely specified.
But the key feature is that they give an abstract description of a system, to varying degrees of specificity,
without grounding in any other ontology.
The only criteria necessary is that the system must be able to be interpreted as following the prescribed computation,
and the conclusion regarding the consciousness or unconsciousness of the system follows.

Below is a quick survey of existing computational accounts of consciousness.
Although my perspective is not convinced of these theories as full accounts of conscious pheonomena, I believe
there are key lessons to learn from each of these systems:

\begin{enumerate}
  \item \textbf{Gödel, Escher, Bach (GEB):} In the his classic book\footnote{\cite{hofstadter1979}, explored further in \cite{hofstadter2007}.}, Douglas Hofstadter speculates that consciousness is an emergent property that arises in
    complex, self-referential systems.
  \item \textbf{Global Workspace Theory (GWT):} In GWT\footnote{\cite{baars2005}}, it is proposed that the brain processes information in separate
    modules; when these modules broadcast and integrate their information in the ``global workspace'', the system becomes
    conscious.\footnote{Though the general approach to GWT is explicitly or implicitly computational, some approaches are grounded in a specific ontology, such as \cite{dehaene2011}.}
  \item \textbf{Attention Schema Theory (AST):} Michael Graziano\footnote{\cite{graziano2015}, expanded upon in \cite{graziano2019}.} has proposed that consciousness arises when a system produces
    an attention model.
    Simpler systems may have self-models and may have attention.
    Consciousness arises when a system gains a model of \emph{its own} attention.
  \item \textbf{Integrated Information Theory (IIT):} In IIT\footnote{\cite{tononi2016}}, physical systems are analyzed in terms of $\Phi$, a measure
    of the causal interdependence between the parts of a system.
    The higher the value of $\Phi$, the more conscious a system is determined to be.
\end{enumerate}

\subsection{The Ontology of Conscious Experience}\label{subsec:ontology}

The thing we can be most sure of is the existence of our own subjective experience.\footnote{This has been clear since at least \cite{descartes1984}.}
Whatever ways we may otherwise be deceived, the state of being deceived itself requires it.
In this paper, I take it that our subjective experience exists,
and that it holds qualitative properties (what philosophers often call qualia).

I call them ``qualitative properties'' (as opposed to qualia) because it comes without additional connotations.
In absence of clear evidence, I prefer to avoid terms that might convey specific assumptions about the nature of these
qualitative phenomena.
These qualitative properties consist in the ineffable character of experiences themselves—the smell of coffee or
sensation of pain.

Roughly speaking, the qualitative properties seem to hold close ties to relational and structural properties,
as they themselves may hold relations and can be structured.
But their qualitative nature itself is not describable only in terms of relationships and structures.

No description will ever convey an experience that has no analogue in the reader's previous experience.
To illustrate, if I tell a normal-sighted person to imagine a color between two shades of blue they are already familiar
with—say cobalt blue and royal blue—they will have no trouble imagining the color.\footnote{This example is drawn from \cite{hume1739}.}
In their mind, it will be almost as if they were presently perceiving the new shade itself.
Cued by the prompts ``cobalt'' and ``royal'', they draw upon their previous perceptions to produce new images.
But a person blind from birth is unable not only to imagine any shade of the color blue, but neither the concept of color
itself, nor even of visual extension.

We might pause to ask the question: Why this limitation?
What is it about a linguistic description that is unable to capture the qualitative properties of an experience?

One difficulty is due to the abstract nature of language.
Words can refer indexically to concepts already held.
They can spell out new relationships between existing concepts, which new relationships may then become novel concepts themselves.
These concepts conjure mental images or sensations based on perceptions we associate with them.
But they cannot create new qualitative categories of experience that are not already available for the mind to manipulate.

Qualitative experiential properties are \emph{ontological} in nature.
This is important: \emph{Qualitative properties are fixed to the systems they are embodied by}.
Any qualitative property is available only to the thing which it is.

My experiences are my own.
I can never share your exact experiences without \emph{being} you.

Though I may sometimes be capable of mimicking some of your relevant state within myself.
We can do this imperfectly by manipulating our experience, say by looking at the same painting.
We can induce similar physical changes to the systems which embody our experience themselves (most saliently our brains),
for instance by sharing in the same psychoactive substances.
We can also communicate linguistic descriptions about items in our common experience, which are then reconsituted in
the mind of the receiver, as we have briefly discussed.

\subsection{The Epistemology of Conscious Experience}\label{subsec:epistemology}

If we were to approach some foreign system and ask whether it has conscious experiences similar to our own, we have two
distinct epistemological avenues to provide evidence in favor of the question.

The first approach is to look at the makeup of the examined system, and compare it to one which we already know is associated with
qualitative experience, most saliently our own (which I'll refer to as the target system).
If the makeup of the examined system is similar to the target system—both in structure and in substance, and especially at
multiple scales of observation—we can gain credence that the experience of the examined system is similar to that of the target system.
This approach is still prone to failure, however.
Without the right devices and expertise, a sleeping brain will look very much like a waking one.

The second approach is to look at the behavior of the examined system, and compare it to the behaviors of the target system
we know are associated with specific experiences.
For instance, I know when I experience hunger, I may fidget while waiting in a restauarant.
If I examine this behavior in another person, I can infer that they may be hungry.
This approach also is subject to its own limitations.
An LLM may mimic human linguistic behavior fairly convincingly.
It may even claim that it is hungry, while we know that there is no sense in which it could actually be hungry.

Each approach alone is fairly weak, but we can gain strong evidence when these two avenues coincide.
When I see my friend fidgeting in a restaurant, I not only have their familiar behavior to show as evidence.
I also know that ontologically we embody extremely similar systems at every level from atoms, to molecules, to
macromolecules, to organelles, to cells, to tissues, to organs, all the way up to beings as a whole.
If a loud noise occurs unexpectedly, the physical chain of events from vibrations of air molecules in sounds waves,
through the vibrations of anatomical parts within our ears, through the stimulation of similar brain structures,
then results in similar behavior—a jump or quick shriek, from both of us.

This conjunction of similarities in the makeup of each system, along with the similarities in behavior, provide
very strong evidence that my friend is conscious in much the same sense that I am.
One who claimed they were \emph{not} conscious would have a burden to provide evidence of similar or greater strength.

Octopodes are similar to us on the smallest levels of organization, but they have a differently-organized central nervous
system and body plan.
Their behavioral patterns are vaguely familiar yet alien.
This evidence indicates that they likely have some qualitative experience analogous to our own, but that its specific
qualitative character is likely quite different.

In a machine learning model, which has almost no resemblance in physical character to that of a human being, it's
highly speculative (I would even say doubtful) that its qualitative properties, if any, are likely to resemble our
own.
The primary argument in its favor is to claim that the model is similar to human beings on an abstract ``computational''
level.

\subsection{The Limitations of Abstractions}\label{subsec:abstractions}

Computations are abstractions.
They express patterns of representations and transformations independent of their implementation within some substrate.
Abstract and functional thinking are critical to our ability to operate usefully within the world.
They allow us to partition out representations from the world and manipulate them in ways that are useful to us,
without needing to grasp an intractable set of details.

The essential question is this:

\begin{quote}
  \emph{Does the substantive implementation of a system matter with respect to
  conscious experience?
  Or can you dispose of the implementation, choose an arbitrary level of description of a system, and give a full
  ontological account of conscious phenomena without reference to its substrate?}
\end{quote}

In my mind, the evidence we have points firmly in the direction that the implementation matters.
We can see at least two fundamental limitations of any abstract description, including computational, functional,
formal, and linguistic descriptions.

First, any abstract description of a system is devoid of qualitative properties.
It may \emph{reference} qualitative properties, but we have already examined that those references can only serve to
spur an existing sytem to reconsistitute qualitiative properties that are already available to it.
A description is  not the same as the thing described.
Some specified computation, an algorithm $X$, is a \emph{description} of the systems which implement it.
It's important that we don't conflate between the map and the territory.

Second, to manifest the qualitative properties of a system, it appears to be a requirement that one must \emph{be} that
system (or at least mimic its substance and structure closely).
But concrete existence is inherently unavailable to abstractions, which by their nature, \emph{discard} their
implementation.
Even if two concrete systems have identical substance and structure, they still implement two \emph{separate}
qualitative experiences.
These properties are not abstract, by the very nature of what it means to be abstract.

We will see in the following sections some additional formal and conceptual challenges of describing concrete systems
computationally.
The purpose of outlining these challenges is so that we can approach them with the goal to meet them.

One of the key challenges is how we can determine which computations are valid descriptions of which systems.
Multiple functional descriptions can be provided for any concrete system, and multiple concrete implementations can
be provided for any functional description.

The sections below will allow us to build a framework for expressing these challenges.
We will put the same framework to use later on to make progress in overcoming them.

\subsection{Describing Systems Computationally}\label{subsec:describing}

Before we proceed further, we will profit to explore what it means to describe a system computationally.
We consider the set of systems $\mathbf{S}$ and the set of computational descriptions $\mathbf{C}$.
System $S \in \mathbf{S}$ implements a computational description $C \in \mathbf{C}$ if and only if there exist mappings $\phi$ and $\gamma$ such that:

\begin{enumerate}
  \item $\phi : \mathbf{S} \to \{s_0, s_1, \ldots \}$ is a mapping from $\mathbf{S}$ to a discrete set of representations $\{s_0, s_1, \ldots \}$.
  It can be useful to denote $\phi(\tilde{S}) \subseteq \{s_0, s_1, \ldots \}$ as the subset of representations for all valid states of the ontological system $S$.
  The states $\{c_0, c_1, \ldots \}$ of $C$ are given within its description.
  \item $\gamma : \{s_0, s_1, \ldots\} \to \{c_0, c_1, \ldots\}$ is a structure-preserving map from the representations of $S$ to the states of $C$.
  \item For every possible state $c_i$ of $C$, there exists a corresponding representation $s_i$ of $\phi(S)$ such that $\gamma(s_i) = c_i$.
  \item For every state transition $(c_i \to c_j)$ in $C$, if $c_i = \gamma(s_i)$, then there exists a valid transition between representations $(s_i \to s_j)$ such that $c_j = \gamma(s_j)$.
  \item $\gamma$ is compatible with $\phi$, meaning that for any two systems $S$ and $S'$, if $\phi(S) = \phi(S')$, then $[\gamma \circ \phi](S) = [\gamma \circ \phi](S')$.
  \item Pragmatically, the mappings $\phi$ and $\gamma$ must be robust under small perturbations of $S$ in some sense we'll leave only intuitively defined, ensuring that the implementation is stable and not overly sensitive to minor fluctuations.
\end{enumerate}

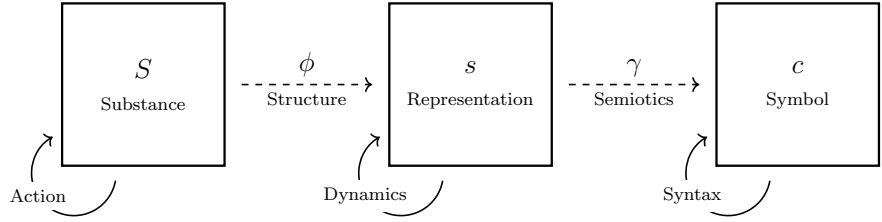
\begin{figure}[tb]
  \centering
  \resizebox{.85\textwidth}{!}{
    \begin{tikzpicture}[
    element/.style={
      draw,
      minimum size=2.5cm,  % Increased size to accommodate two lines of text
      inner sep=5pt,
      align=center,
      line width=1pt,
      font=\large
    },
    arrow/.style={->, shorten >= 7pt, shorten <= 7pt, line width=.7pt},
    mainlabel/.style={font=\large},
    label/.style={font=\footnotesize}  % Even smaller font for the labels
    ]
    % Nodes (squares) with names and labels inside
    \node[element] (e0) {$S$ \\ \footnotesize Substance};
    \node[element, right=2.5cm of e0] (e1) {$s$ \\ \footnotesize Representation};
    \node[element, right=2.5cm of e1] (e2) {$c$ \\ \footnotesize Symbol};

    % Arrows with labels
    \draw[arrow,dashed] (e0) -- (e1)
    node[mainlabel, midway, above] {$\phi$}
    node[label, midway, below] {Structure};
    \draw[arrow,dashed] (e1) -- (e2)
    node[mainlabel, midway, above] {$\gamma$}
    node[label, midway, below] {Semiotics};

    \drawloop[<-,label,stretch=1.1,line width=.7pt]{e0}{210}{260} node[pos=.4,fill=white]{Action};
    \drawloop[<-,label,stretch=1.1,line width=.7pt]{e1}{210}{260} node[pos=.4,fill=white]{Dynamics};
    \drawloop[<-,label,stretch=1.1,line width=.7pt]{e2}{210}{260} node[pos=.4,fill=white]{Syntax};
    \end{tikzpicture}
  }
  \caption{In this diagram, we see the interplay of the system's substance $S$, its structural representation $\phi(S) = s$,
    and its semiotic interpretation $\gamma(s) = c$.}
  \label{fig:computation}
\end{figure}

In this formulation, $\phi$ can be seen as the system's \emph{structure}, which produces a state representation $s = \phi(S)$.
The ability for the system to assume these representations are what make the system amenable to computation,
whose transitions $(s_i \to s_j)$ are the representation's \emph{dynamics}.

$\gamma$ is the system's \emph{semiotics}, the assignment of a symbolic state $c = \gamma(s)$ to the representation of the system.
The symbolic states evolve according to state transitions $(c_i \to c_j)$ which are its \emph{syntax}.
The evolution of the system's dynamics must match the computational description's syntax in order for $S$ to implement $C$
via $[\gamma \circ \phi]$.

The set of state transitions in $C$ can be represented as a matrix $T_C$ where each $t_{ij} \in T_C$ is 1 when $(c_i \to c_j)$ is a valid transition and 0 otherwise.
A corresponding matrix $T_{\phi(\tilde{S})}$ exists for the representations $\phi(\tilde{S})$.

$\Theta : \mathbf{S} \to 2^{\mathbf{C}}$ is a map of \emph{interpretations} from each system $S \in \mathbf{S}$ into its
set of possible computational descriptions, a subset of $\mathbf{C}$.
Each $\Theta(S)$ can be seen as a composition $[\gamma \circ \phi]$ of some chosen $\gamma$ and $\phi$, which must meet the
compatibility condition described above.

Conversely, $\Omega : \mathbf{C} \to 2^{\mathbf{S}}$ is a map of \emph{implementations} from each computational description $C \in \mathbf{C}$ into its
set of possible realized systems, a subset of $\mathbf{S}$.
Each $\Omega(C)$ can be seen as a composition $[\phi^{-1} \circ \gamma^{-1}]$ (note the reversal of order), the inverse relations of some chosen $\gamma$ and $\phi$.
Note that each $\phi^{-1}$ and $\gamma^{-1}$  may be nondeterministic mappings (multivalued inverse functions) which may assign multiple possible system states to a given
computational state.

\subsection{Computational Ambiguity}\label{subsec:ambiguity}

The nature of an abstraction is to be ambiguous.\footnote{These ambiguities have also been explored in very interesting fashion in~\cite{shagrir}.
The ``Physical Realization Theorem'' explored in~\cite{putnam1988} draws a stronger conclusion about computational ambiguity which is contested
  (see~\cite{chalmers1996}).}
Abstractions generalize over concretions to extract meaningful common structures, relationships, and behaviors
between them.

There are two key distinctions we will observe in the types of ambiguities we find between systems and
computational descriptions.

The first is a distinction that relates to the direction of the ambiguity.
We refer to the case that $\Theta(S)$ of some system $S$ has multiple computational descriptions $C_1, C_2, \dots$ as \emph{implementational ambiguity}.
We refer to the alternate case, where $\Omega(C)$ of some computational description $C$ may be implemented as multiple systems $S_1, S_2, \dots$, as \emph{interpretational ambiguity}.

The second distinction is between \emph{structural} and \emph{semiotic ambiguity}, which result from the maps $\phi$ and $\gamma$, respectively.

\subsection{Implementational Ambiguity}\label{subsec:implementational-ambiguity}

Implementational ambiguity occurs between a computational description $C$ and multiple concrete implementations $S_1, S_2, \dots$.
This is commonplace, and can be seen in our ability to run the same computer programs on multiple machines.
Implementational ambiguity allows us to describe disparate physical systems in terms of the same functional account, and therefore
highlight functional similarities between those systems.

Implementational ambiguity is not problematic for a computational ontology of conscious experience because of the ambiguity on
its own.
However, when the makeup of those systems is vastly different, it could be a vulnerable assumption
that such disparate systems share qualitative properties without sharing similarity in substance or structure.

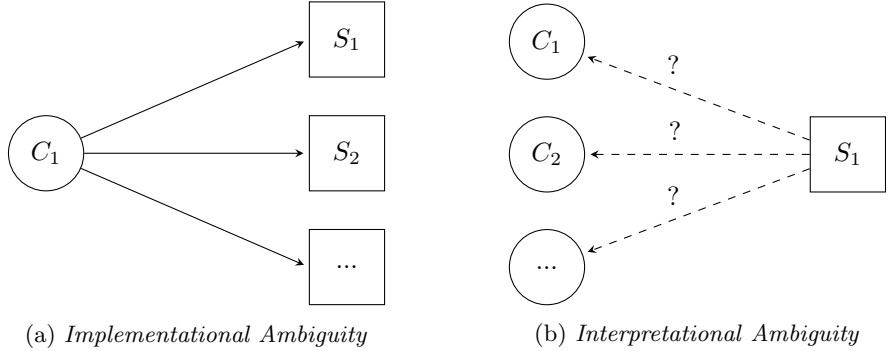
\begin{figure}[t]
  \centering
  \begin{subfigure}[b]{0.45\textwidth}
    \centering
    \begin{tikzpicture}[
    box/.style={draw, rectangle, minimum width=1cm, minimum height=1cm},
    arrow/.style={->, >=stealth, shorten >= 2pt}
    ]

% Boxes on the right
    \node[box] (S1) at (4,5.5) {$S_1$};
    \node[box, below=0.5cm of S1] (S2) {$S_2$};
    \node[box, below=0.5cm of S2] (S3) {...};

% Calculate the vertical center of the boxes
    \path let \p1 = (S1.north), \p2 = (S3.south) in
    coordinate (center) at (0, {(\y1 + \y2)/2});

% Circle on the left, centered vertically
    \node[draw, circle, minimum size=1cm] (C1) at (center) {$C_1$};

% Arrows from circle to boxes
    \draw[arrow] (C1) -- (S1.west);
    \draw[arrow] (C1) -- (S2.west);
    \draw[arrow] (C1) -- (S3.west);

    \end{tikzpicture}
    \caption{\emph{Implementational Ambiguity}}
    \label{fig:diagram1}
  \end{subfigure}
  \hspace{.01\textwidth}
  \begin{subfigure}[b]{0.45\textwidth}
    \centering
    \begin{tikzpicture}[
    arrow/.style={->, >=stealth, dashed, shorten >= 2pt}
    ]

% Circles on the left
    \node[draw, circle, minimum size=1cm] (C1) at (0,1.5) {$C_1$};
    \node[draw, circle, minimum size=1cm] (C2) at (0,0) {$C_2$};
    \node[draw, circle, minimum size=1cm] (C3) at (0,-1.5) {...};

% Box on the right
    \node[draw, rectangle, minimum width=1cm, minimum height=1cm] (S1) at (4,0) {$S_1$};

% Arrows from circles to box
    \draw[arrow] (S1) -- (C1) node[label, pos=0.6, above=2pt] {?};
    \draw[arrow] (S1) -- (C2) node[label, pos=0.6, above=2pt] {?};
    \draw[arrow] (S1) -- (C3) node[label, pos=0.6, above=2pt] {?};

    \end{tikzpicture}
    \caption{\emph{Interpretational Ambiguity}}
    \label{fig:diagram2}
  \end{subfigure}
  \caption{\emph{Implementational ambiguity} expresses the ability of a single computation $C_1$ to be implemented in one of many
  physical systems, $S_1, S_2, \ldots$ \
  \emph{Interpretational ambiguity}, on the other hand, points out our inability to give a unique computational interpretation
    of the behavior of any given physical system, shown in the diagram with a single physical system $S_1$ able to be accounted for by
  multiple computational systems $C_1, C_2, \ldots$}
  \label{fig:ambiguity}
\end{figure}
\subsection{Interpretational Ambiguity}\label{subsec:interpretational-ambiguity}

\emph{Interpretational ambiguity} runs in the opposite direction, \emph{from a physical implementation to multiple computational descriptions}.
But if there are multiple computations that serve as descriptions of the same concrete system, how do we know which is
the correct one?
We'll consider questions surrounding the mapping between systems and computational descriptions below.

\subsection{Semiotic Ambiguity}\label{subsec:semiotic-ambiguity}

Semiotic ambiguity occurs in the relation specified by $\gamma$, the mapping between a system state $s = \phi(S)$ and
its corresponding computational state $c$.
There is a set of relations $V$ where each $\gamma \in V$ is a distinct mapping between structural representations
and symbolic states.
Our choice between potentially multiple possible $\gamma \in V$
produces the ambiguity.

Let's illustrate the ambiguity with a simple example that frequently occurs in your own computer
hardware.
In the example provides below we take our system $S$, with each state $\phi(S)$ consisting of 6 bytes $w_0, \dots, w_5$,
and each byte containing 7 bits $x_0, \ldots, x_6$.
We will show that a single initial state $s_0$ and transition $\delta$ to the resulting state $s_1 = \delta(s_0)$
of this system can be mapped to a computational description in no less than four intuitive (and largely realistic) ways.

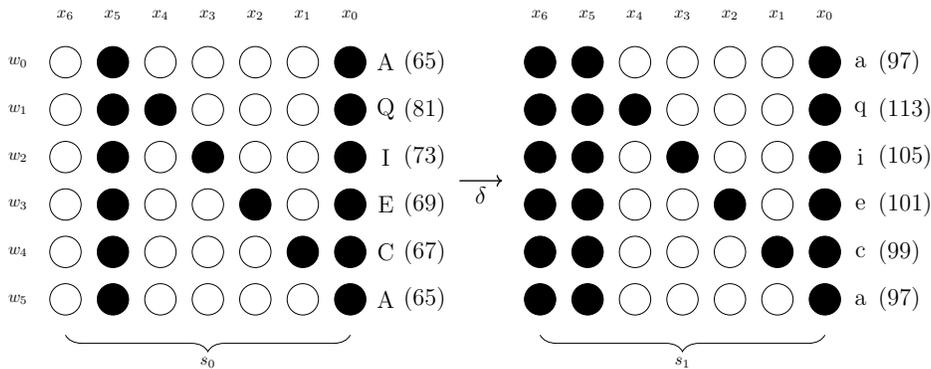
\begin{figure}[h]
  \centering
  \resizebox{.9\textwidth}{!}{
    \centering
    \begin{tikzpicture}
      \node at (0,5) {$w_0$};
      \node at (0,4) {$w_1$};
      \node at (0,3) {$w_2$};
      \node at (0,2) {$w_3$};
      \node at (0,1) {$w_4$};
      \node at (0,0) {$w_5$};

      \node at (1,6) {$x_6$};
      \node at (2,6) {$x_5$};
      \node at (3,6) {$x_4$};
      \node at (4,6) {$x_3$};
      \node at (5,6) {$x_2$};
      \node at (6,6) {$x_1$};
      \node at (7,6) {$x_0$};

      \draw[fill=white] (1, 5) circle (.33cm);
      \draw[fill=black] (2, 5) circle (.33cm);
      \draw[fill=white] (3, 5) circle (.33cm);
      \draw[fill=white] (4, 5) circle (.33cm);
      \draw[fill=white] (5,5) circle (.33cm);
      \draw[fill=white] (6,5) circle (.33cm);
      \draw[fill=black] (7,5) circle (.33cm);
      \node[font=\Large] at (7.75,5) {A};
      \node[anchor=west,font=\Large] at (8,5) {(65)};

      \draw[fill=white] (1, 4) circle (.33cm);
      \draw[fill=black] (2, 4) circle (.33cm);
      \draw[fill=black] (3, 4) circle (.33cm);
      \draw[fill=white] (4, 4) circle (.33cm);
      \draw[fill=white] (5,4) circle (.33cm);
      \draw[fill=white] (6,4) circle (.33cm);
      \draw[fill=black] (7,4) circle (.33cm);
      \node[font=\Large] at (7.75,4) {Q};
      \node[anchor=west,font=\Large] at (8,4) {(81)};

      \draw[fill=white] (1, 3) circle (.33cm);
      \draw[fill=black] (2, 3) circle (.33cm);
      \draw[fill=white] (3, 3) circle (.33cm);
      \draw[fill=black] (4, 3) circle (.33cm);
      \draw[fill=white] (5,3) circle (.33cm);
      \draw[fill=white] (6,3) circle (.33cm);
      \draw[fill=black] (7,3) circle (.33cm);
      \node[font=\Large] at (7.75,3) {I};
      \node[anchor=west,font=\Large] at (8,3) {(73)};

      \draw[fill=white] (1, 2) circle (.33cm);
      \draw[fill=black] (2, 2) circle (.33cm);
      \draw[fill=white] (3, 2) circle (.33cm);
      \draw[fill=white] (4, 2) circle (.33cm);
      \draw[fill=black] (5,2) circle (.33cm);
      \draw[fill=white] (6,2) circle (.33cm);
      \draw[fill=black] (7,2) circle (.33cm);
      \node[font=\Large] at (7.75,2) {E};
      \node[anchor=west,font=\Large] at (8,2) {(69)};

      \draw[fill=white] (1, 1) circle (.33cm);
      \draw[fill=black] (2, 1) circle (.33cm);
      \draw[fill=white] (3, 1) circle (.33cm);
      \draw[fill=white] (4, 1) circle (.33cm);
      \draw[fill=white] (5,1) circle (.33cm);
      \draw[fill=black] (6,1) circle (.33cm);
      \draw[fill=black] (7,1) circle (.33cm);
      \node[font=\Large] at (7.75,1) {C};
      \node[anchor=west,font=\Large] at (8,1) {(67)};

      \draw[fill=white] (1, 0) circle (.33cm);
      \draw[fill=black] (2, 0) circle (.33cm);
      \draw[fill=white] (3, 0) circle (.33cm);
      \draw[fill=white] (4, 0) circle (.33cm);
      \draw[fill=white] (5,0) circle (.33cm);
      \draw[fill=white] (6,0) circle (.33cm);
      \draw[fill=black] (7,0) circle (.33cm);
      \node[font=\Large] at (7.75,0) {A};
      \node[anchor=west,font=\Large] at (8,0) {(65)};

      \draw[decorate,decoration={brace,mirror,raise=10pt,amplitude=10pt}] (1, -.4) -- (7, -.4) node [midway,below=20pt] {$s_0$};

      \draw[->, thick] (9.3, 2.5) -- (10.2,2.5) node[below, midway, font=\Large] {$\delta$};

      \node at (11,6) {$x_6$};
      \node at (12,6) {$x_5$};
      \node at (13,6) {$x_4$};
      \node at (14,6) {$x_3$};
      \node at (15,6) {$x_2$};
      \node at (16,6) {$x_1$};
      \node at (17,6) {$x_0$};

      \draw[fill=black] (11, 5) circle (.33cm);
      \draw[fill=black] (12, 5) circle (.33cm);
      \draw[fill=white] (13, 5) circle (.33cm);
      \draw[fill=white] (14, 5) circle (.33cm);
      \draw[fill=white] (15,5) circle (.33cm);
      \draw[fill=white] (16,5) circle (.33cm);
      \draw[fill=black] (17,5) circle (.33cm);
      \node[font=\Large] at (17.75,5) {a};
      \node[anchor=west,font=\Large] at (18,5) {(97)};

      \draw[fill=black] (11, 4) circle (.33cm);
      \draw[fill=black] (12, 4) circle (.33cm);
      \draw[fill=black] (13, 4) circle (.33cm);
      \draw[fill=white] (14, 4) circle (.33cm);
      \draw[fill=white] (15,4) circle (.33cm);
      \draw[fill=white] (16,4) circle (.33cm);
      \draw[fill=black] (17,4) circle (.33cm);
      \node[font=\Large] at (17.75,4) {q};
      \node[anchor=west,font=\Large] at (18,4) {(113)};

      \draw[fill=black] (11, 3) circle (.33cm);
      \draw[fill=black] (12, 3) circle (.33cm);
      \draw[fill=white] (13, 3) circle (.33cm);
      \draw[fill=black] (14, 3) circle (.33cm);
      \draw[fill=white] (15,3) circle (.33cm);
      \draw[fill=white] (16,3) circle (.33cm);
      \draw[fill=black] (17,3) circle (.33cm);
      \node[font=\Large] at (17.75,3) {i};
      \node[anchor=west,font=\Large] at (18,3) {(105)};

      \draw[fill=black] (11, 2) circle (.33cm);
      \draw[fill=black] (12, 2) circle (.33cm);
      \draw[fill=white] (13, 2) circle (.33cm);
      \draw[fill=white] (14, 2) circle (.33cm);
      \draw[fill=black] (15,2) circle (.33cm);
      \draw[fill=white] (16,2) circle (.33cm);
      \draw[fill=black] (17,2) circle (.33cm);
      \node[font=\Large] at (17.75,2) {e};
      \node[anchor=west,font=\Large] at (18,2) {(101)};

      \draw[fill=black] (11, 1) circle (.33cm);
      \draw[fill=black] (12, 1) circle (.33cm);
      \draw[fill=white] (13, 1) circle (.33cm);
      \draw[fill=white] (14, 1) circle (.33cm);
      \draw[fill=white] (15,1) circle (.33cm);
      \draw[fill=black] (16,1) circle (.33cm);
      \draw[fill=black] (17,1) circle (.33cm);
      \node[font=\Large] at (17.75,1) {c};
      \node[anchor=west,font=\Large] at (18,1) {(99)};

      \draw[fill=black] (11, 0) circle (.33cm);
      \draw[fill=black] (12, 0) circle (.33cm);
      \draw[fill=white] (13, 0) circle (.33cm);
      \draw[fill=white] (14, 0) circle (.33cm);
      \draw[fill=white] (15,0) circle (.33cm);
      \draw[fill=white] (16,0) circle (.33cm);
      \draw[fill=black] (17,0) circle (.33cm);
      \node[font=\Large] at (17.75,0) {a};
      \node[anchor=west,font=\Large] at (18,0) {(97)};

      \draw[decorate,decoration={brace,mirror,raise=10pt,amplitude=10pt}] (11, -.4) -- (17, -.4) node [midway,below=20pt] {$s_1$};
    \end{tikzpicture}
  }
  \caption{The operation $\delta$, a transition from state $s_0$ to $s_1$, can be understood as no less than four different computations.
  Note that each circle indicates a single bit, with a black circle indicating the 1 state and a white circle indicating the 0 state.}
  \label{fig:ambigious}
\end{figure}

\subsubsection{Pixel Semiotics}

The first interpretation of the system will be a visual one, with each bit $x_i$ representing a pixel to be
drawn to the screen, and each byte $w_i$ representing the next row.
Our example shows a pattern of pixels that could be interpreted as the letter `N'.
In this case, the transition $\delta$ represents an operation to draw a vertical line in the far left column,
making the letter bold.

We might express this transformation in some computational description $C$ as
\[ \text{N} \xrightarrow[\text{bold}]{} \mathbf{N}, \]
where $\text{N}$ and $\mathbf{N}$ are states of $C$ and $\xrightarrow[\text{bold}]{}$ is $\delta$, the state transition.

\subsubsection{Alphabetic Semiotics}

We may also take an alphabetic interpretation, in which each byte $w_i$ represents a letter of the English alphabet.
In this example, we're used the ASCII encoding, which conveniently is defined so that the number that encodes any
uppercase letter plus 32 becomes the ASCII encoding of the same letter in lowercase.
In other words, the application of $\delta$ makes the letters lowercase.

\[ \text{AQIECA} \xrightarrow[\text{lowercase}]{} \text{aquieca} \]

There is a further ambiguity in this example in that if we used a different encoding, each number would represent a
different character, and the addition of 32 to each encoded character is not likely to have any neat semiotic
interpretation.

\subsubsection{Numeric Semiotics}

We can also interpret the system numerically, with each byte $w_i$ representing a binary integer.
In this case, the application of $\delta$ increments each number in the sequence by 32.
\[ 65 \xrightarrow[+32]{} 97 \]
\[ 81 \xrightarrow[+32]{} 113 \]
\[ \dots \]

\subsubsection{Bitmask Semiotics}

A bitmask takes a binary representation at face value, i.e.\ as a sequence of independent bits.
They are often used to represent a series of boolean (i.e.\ true=1/false=0) variables within a single byte.
In our example, each $x_i$ would represent some true/false value.
The computation expressed by $\delta$ can then be interpreted as a toggling on of the $x_6$ bit in each byte.
In this example we'll examine the operation $\delta$ on $w_0$, which can be seen in boolean context as
\[ \text{FTFFFFT} \xrightarrow[x_6=\text{T}]{} \text{TTFFFFT}, \]
or as a programmer might express it equivalently using bitwise operations
\[ \text{0100001} \ | \ \text{1000000} \ = \ \text{1100001}, \]
where the symbol `|' indicates the bitwise ``or'' operation, which produces a resulting 1 in some position $x_i$ if
\emph{either} of the operands' $x_i$ is also a 1.

\subsection{Structural Ambiguity}\label{subsec:implementation-ambiguity}

\emph{Structural ambiguity} is found in the mapping $\phi(S)$ which ``slices up'' a system $S$ into discrete
representations and state transitions.
The property of a system having a set of $U = \{\phi_a, \phi_b, \ldots\}$ where some $\phi \in U$ are useful
implementations of computational descriptions is what makes them amenable as computing machines.

For instance, the implementation of a computer's memory as an array of binary (off or on) digits is itself an abstraction
over the actual physical system.
In an actual computer $S$, each bit in memory is represented by an electric potential.
$\phi$ maps the system $S$ into a useful structural representation via $\phi(S)$.

To illustrate, let's say that 0V represents the 0 or ``off'' state and 1V represents the 1 or ``on'' state.
At any point in time, the likelihood of the component that represents a given bit actually being \emph{equal to} 0V or
1V is extremely small.
Instead, the computer is engineered to behave as ``on'' if the voltage is above a certain threshold $\tau$ (say $\tau = .5\text{V}$) and
``off'' if the voltage falls below the threshold.
Given our computer $S$, we can isolate some component $x$ as a subsystem.
\[ \phi(x < \tau) = 0 \text{ and } \ \phi(x \geq \tau) = 1 \]

When $\tau = .5\text{V}$, the subsystem will have an identical interpretation if
\[\phi(x=.51\text{V}) = \phi(x=.99\text{V}) = 1. \]

This example of implementational ambiguity can seem fairly mild in the case of systems with similar architecture, but result
in comparisons of vastly different systems in actual practice (comparisons between engineered and biogical systems,
for instance).

Even the binary interpretation we require for the computer to be useful needs to be engineered onto and read out from
the physical system.
Without changing the engineering of the memory or logical units of a machine (though to be practical you'd need
to update the input-output systems), you could alter the threshold to some arbitrary other voltage (say
$\tau = .25\text{V}$) and have a vastly different interpretation of the computations that have occured in the system.
Even further, without making any changes to the the threshold, you could simply reverse the interpretation as
$\phi'(x) = \neg\phi(x)$, with untold havoc in the results.
\[ \phi'(x < \tau) = 1 \text{ and } \phi'(x \geq \tau) = 0 \]

\subsection{Ontological Inconsistency}\label{subsec:inconsistency}

While we've already viewed the results of computational abiguity in more relatable settings in the examples above,
this section will take the concept further.

We will assume that any closed physical system can be approximated to arbitrary accuracy as a finite automata.
This assumption is evidenced by various known methods for approximating continuous phenomena with discrete methods
such as Taylor's Theorem, Fourier Series, and the Shannon-Nyquist
Sampling Theorem,\footnote{\cite{shannon1949}} as well as techniques of art in computational physics.
Given that assumption, we can demonstrate that:

\begin{enumerate}
  \item There always exist multiple computational descriptions consistent with any closed physical system; and that
  \item A pair of those descriptions will be mutually inconsistent.
\end{enumerate}

We will consider a physical system described by a finite automata
$\mathbf{Q}$, where the state of the machine consists of a
sequence of symbols $s$ of finite length $N$, where the $i$-th symbol, $i < N$, in $s$ is represented $s_i \in \{0,1\}$)
and a set of transition rules $R$, where each $r \in R$ is a rule
\[r(s) = s',\]
which determines a subsequent
state $s'$ at time $t+1$ according to the current state $s$ at time $t$.
Each $r$ is composed of a series of transitions of individual symbols within the state
\[r(s_i) = s_i'.\]

Any such system $\mathbf{Q}$ is consistent with a more general system $\mathbf{Q'}$ where each $r_i'(s)$ is defined
\[r'(s_{i+kN}) = r(s_i),\]
where $k$ is an integer $k \geq 0$ and $i < N$.\footnote{Note that $\mathbf{Q'}$ is no longer a finite machine, which is a
requirement only for our base description $\mathbf{Q}$ of a the physical system.}
This generalizes $r$ outside the range ${0 \leq i < N}$.
This property \emph{guarantees} that all such finite systems will be interpretationally and semiotically ambiguous,
since both $\mathbf{Q}$ and its extended $\mathbf{Q'}$ are correct models of the system.

However, any such system $\mathbf{Q}$ is \emph{also} consistent with a system $\mathbf{Q''}$ where each $r''(s_j)$ is defined
\[
  r''(s_j) = \begin{cases}
    r'(s_j) & \text{if } j < N \\
    1 & \text{if } j \geq N \text{ and } r'(s_j) = 0 \\
    0 & \text{if } j \geq N \text{ and } r'(s_j) = 1.
  \end{cases}
\]
Note that this is a form of diagonalization, where $r''(s_j)$ just produces the opposite of $r'(s_j)$ when $j \geq N$,
thus ensuring its inconsistency.
This argument bears no accidental relation to
Cantor's diagonal argument, Russell's paradox,
Gödel's incompleteness theorems, and Turing's halting problem.\footnote{\cite{cantor1891}, \cite{russell1903}, \cite{godel1992}, \cite{turing1937}}
It also is due some credit to an argument made by Wittgenstein regarding arithmetical sequences.\footnote{\cite{wittgenstein1953}}

While $\mathbf{Q}$ is consistent with both $\mathbf{Q'}$ and $\mathbf{Q''}$, $\mathbf{Q'}$ and $\mathbf{Q''}$ are
\emph{mutually inconsistent} with each other when $j \geq N$
Expressed formally,
\[ \mathbf{Q'} \models \mathbf{Q} \text{ and } \mathbf{Q''} \models \mathbf{Q}, \]
but
\[ \mathbf{Q'} \models \neg\mathbf{Q''} \text{ and } \mathbf{Q''} \models \neg\mathbf{Q'}. \]

\subsection{Mapping Computational Systems}\label{subsec:mapping-computational-systems}

In this section, we'll examine some of the questions that come about when trying to map systems to computational
descriptions, and how we might address them.

\vspace{.5em}
\begin{quote}
  \textbf{\emph{If there are multiple valid but inconsistent computational descriptions of any system, how do we determine which
  description of a system is the correct one?}}
\end{quote}

We might observe that there is no proposition to admit arbitrary computational descriptions into our ontology;
we are only examining systems for the presence of a \emph{specific} computational description.
We will denote this proposition $R : \mathbf{S} \times \mathbf{C} \to \{0,1\}$, where $R(S,C)$ is 1 when a system $S$
implements the specific computational description $C$ according to the criteria from the
section~\nameref{subsec:describing}, and 0 when it does not.

\vspace{.5em}
\begin{quote}
  \textbf{\emph{Does the proposition R require that R itself is computational?}}
\end{quote}

If $R(S,C)$ is taken as a statement that the ontology of consciousness is fully computational, then it seems to require
a computational procedure that will determine whether the system $S$ matches the proposed computation $C$.

Even the weaker epistemological position that $R(S,C)$ can be \emph{known} solely through computational means
would require by definition that $R(S,C)$ is computational.

\vspace{.5em}
\begin{quote}
  \textbf{\emph{Can R(S,C) be computational, if its input S is a concrete system (as opposed to an abstract one)?}}
\end{quote}

Computation can only operate over abstract inputs to produce abstract outputs.
If $R(S, C)$ is supposed to be determined solely through computational
means, we require an abstract representation of the system $s = \phi(S)$ that is amenable to being interpreted
computationally in the first place.

This process can't bootstrap itself; if you suppose that $\phi$ \emph{itself} is only a computation, then $S$ is
already a computational state.
We've only pushed back the question to some $S = \phi'(S')$.
Ultimately some final reference to the concrete (i.e.\ non-abstract) system $S^*$ must be reached.
Any mapping $\phi^*$ from the concrete system $S^*$ to a representation $s^*$ cannot be solely computational.

\vspace{.5em}
\begin{quote}
  \textbf{\emph{If it's not computational, what is it?}}
\end{quote}

$\phi(S)$ is a structural interpretation of the system $S$.
It depends on an interpreting system $I$, an observer, which is capable of taking concrete percepts and producing
abstract representations.
$I$ is a concrete system.

Note that this is not a statement about the Church-Turing-Deutsch principle, which states that a universal computing
device can simulate every physical process.
The interpreting system $I$ may be itself simulable as $\phi(I)$, produced by an observer $I'$, which interestingly
may be $I$ itself, in which case $I' = I$.\footnote{This thought inevitably brings to mind \cite{vonneumann1966}.}
But the simulation $\phi(I)$ cannot perform $\phi$, as the simulation is already abstract; it has no concrete input
available to it.

\vspace{.5em}
\begin{quote}
  \textbf{\emph{Doesn't that introduce a significant influence of arbitrary interpretation into our ontology, if the determination
  is dependent on an observer?}}
\end{quote}

Yes, I would say that any computational account of the ontology of consciousness does admit interpretation into our ontology.
Computational ontologies systems fix our choice of $\gamma$ but leave $\phi$ open.

An alternative, grounding computation to a specific ontology by also fixing $\phi$, is explored in the following sections.
Note that grounding doesn't settle the choice of $\phi$ \emph{epistemologically}.
That's an empirical question that may not be fully decidable (though we will show methods for chosing optimal choices of $\phi$ based on different criteria).
But the stipulation of having a fixed $\phi$ does remove the ambiguity from our \emph{ontology}.

\subsection{Ontological Context}\label{subsec:context}

We have already seen that $R$ cannot be computational.
But it's interesting to consider if it were.
If $R(S,C)$ were computational, it would require a representation $\phi(S)$ to act on which was computational.

Supposing that $\phi$ \emph{were} computational,
it must be \emph{interpretationally ambiguous}, as all finite automata were shown to be above in
the section~\nameref{subsec:inconsistency}.
There will be more than one $\phi$ such that $\phi(S) = s$, and that some of those $\phi$ will be inconsistent
with each other in other contexts.

Consider what happens if we were to claim that there exists some computational procedure
\[ \mu(\{\phi_a, \phi_b, \ldots\}) \to \phi \]
which will determine a unique $\phi$, based on its computational properties alone (i.e.\ without reference to any
substrate).
Such a procedure cannot distinguish between two inconsistent mappings $\phi_a$ and $\phi_b$ which are
isomorphic but whose symbols are ungrounded.

Recall the structural ambiguity we can introduce simply by inverting the representation\footnote{Refer to
the section~\nameref{subsec:implementation-ambiguity}.}
\[\phi_a(S) = \neg \phi_b(S), \text{ i.e. } \phi_a(S) = 0 \text{ and } \phi_b(S) = 1,\]
where 1 and 0 are arbitrary ungrounded symbols.
There is no way for $\mu$ to distinguish computationally between $\phi_a$ and $\phi_b$, unless the
inputs and outputs are grounded with reference to some context.

And yet we know what kind of effect a reversal in representation could have on the operation of a system.
An example was given above, in which we fixed each representation to a set of physical facts about voltages,
eliminating the ambiguity, as in
\[ \phi_a(x < .5\text{V}) = 0 \text{ and } \phi_a(x \geq .5\text{V}) = 1 \]
while
\[ \phi_b(x < .5\text{V}) = 1 \text{ and } \phi_b(x \geq .5\text{V}) = 0. \]

We could counter by claiming that the isomorphism between the systems allows them to be considered identically.
But both $\phi_a$ and $\phi_b$ could in fact be used by different components $a$ and $b$
within the same system $Q$, in which case we'd need to differentiate between them.
\[ \phi(Q) = \big\{\phi_a(a),\,\phi_b(b)\big\} \]
For instance, if we take $a = b = .7\text{V},$
\[ \phi(Q) = \big\{\phi_a(.7\text{V}) = 1,\ \phi_b(.7\text{V}) = 0\big\}. \]

At this point, we can appeal back to the same claim of isomorphism—that within the system $Q$, $\phi_a$ and $\phi_b$
are distinct, but without the larger context, they are not.
This seems like a reasonable statement from an epistemological perspective, but more difficult ontologically,
where questions about context-dependence come into play.
What would it mean for the intrinsic properties of a system to be context-dependent?

In any case, the question can be resolved straightforwardly with reference to an ontological substrate, as with the
introduction of the inputs .7V, which create an invariance between the systems and allows them to be compared.

\subsection{Grounded Computation}\label{subsec:grounded-computation}

We've now seen several lines of motivation for grounding our computations within some ontological context.
Most significant are the principled reasons we outlined in the first few sections of the paper, but we've seen that
ontological context is also adept at removing ambiguities which otherwise provoke questions.

There are two straightforward steps towards the removal of many ontological ambiguities that arise in mapping
substantial systems to computational descriptions.
We can:
\begin{enumerate}
  \item Choose specific mappings $\phi$ and $\gamma$, instead of leaving their choice open; and we can
  \item Explicitly ground each representation: each $s = \phi(S)$ must be identified with an ontological entity or state.
\end{enumerate}

Note that this resolution in no way ensures any particular scheme's correctness; these are only measures to remove
ambiguity.
We will also see that there is some ambiguity that is unavoidable in principle due to epistemological considerations
(as opposed to ontological ones).
We'll demonstrate how the remaining ambiguity arises, and propose methods for measuring and minimizing it.

Because ${[\gamma \circ \phi]}$ yields a single computational state $c$, no further consideration is needed to resolve
computational ambiguities once $\gamma$ and $\phi$ are specifically chosen.
However, the grounding relation between states of $\phi(S)$ and ontological states needs further explication.

Ultimately, we want to anchor our descriptions to the concrete system by identifying representations with the
substantial states of the system.
This process isn't new, and occurs implicitly in many physical theories.
The level the ontological identity will occur at will depend on the representation of the
ontological system $\phi(S)$.

We want our states $\phi(\tilde{S})$ to follow natural representations of the ontological state of $S$.
This essentially amounts to a coarse-graining procedure in which $\phi(\tilde{S})$ can be interpreted as a set of
equivalence classes over the ontological states of $S$.
One method of determining equivalence classes empirically would be by finding embodied characteristics of systems
which are known to be qualitatively similar.

Let's consider the states of a system $S$, a single transistor which may either be off or on.
In this case, $s_0$ will embody the ``off'' state and $s_1$ will embody the ``on'' state.
So we have
\[ \phi(\tilde{S}) = \{s_0, s_1\}\]
based on some criteria we will not specify.
We could imagine this to follow a voltage scheme as we've already visited.
We will notice here that the transistor itself has many more than two physical states, which are coarse-grained out
by our choice of $\phi$.

Based on our epistemological principles, we've taken it that we can give strong evidence that two systems are qualitatively
similar when their behavior is similar (behavior is defined by $\gamma$),
and their ontological substance and structure is similar (substance and structure are defined by $\phi$).
The similarity of substance can then be indicated by some divergence measure between two systems' representations
$\phi(X)$ and $\phi(Y)$ or sets of valid representations $\phi(\tilde{X})$ and $\phi(\tilde{Y})$.
We will indicate this divergence as
\[\mathbb{D}\big[\phi(X) {\lVert} \phi(Y)\big] \text{ or } \mathbb{D}\big[\phi(\tilde{X}) {\lVert} \phi(\tilde{Y})\big].\]
Further discussion about how this divergence may be calculated will follow in the next section.

Given that we cannot know a system's \emph{ontos} (fundamental ontology) $O(S)$, where ${O(S) \neq \phi(S)}$, this is
only an approximate divergence measure between the systems, subject to infidelities in the representation $\phi$ utilized.
The available choice of representations $\phi$ is fixed by our observed data.
In the case of neuroscience, we may have voxel data from various imaging methods, but no direct data about individual
neurons or subcellular structures.

We may have direct or indirect means of surveying multiple levels of representation.
In these cases, we can introduce a partial order between mappings $\phi_a$ and $\phi_b$ in which
\[ \phi_b \succ \phi_a \text{ iff } \phi_b \text{ supervenes on } \phi_a. \]
Here we understand \emph{supervenes on} to mean that there is a surjective mapping
${\rho : \phi_a(S) \to \phi_b(S)}$ for every system $S$.
Every fine-grained state in $\phi_a(S)$ is accounted for by exactly one coarse-grained state in $\phi_b(S)$,
and each state transition $\big(\phi_a(S) \to \phi_a(S')\big)$ is accounted for by a corresponding transition
$\big(\phi_b(S) \to \phi_b(S')\big)$.
In practice, this supervenience relation itself may be fuzzier than indicated here.
We can use divergence metrics that measure to what extent one representation is accounted for by the other, with discussion
to follow in the following sections.

Then for known $\phi_a$ and $\phi_b$, such that $\phi_b \succ \phi_a$, we can place a lower bound on the divergence from \emph{ontos}
as
\[ \mathbb{D}\big[\phi_b(S) {\lVert} O(S)\big] \ \geq \ \mathbb{D}\big[\phi_b(S) {\lVert} \phi_a(S)\big], \]
giving an indication of how much fidelity we lost by introducing our coarse-graining.
Depending on the choice of divergence measure, this inequality can be proven using the Data Processing Inequality
due to the Markov chain
\[
O(S) \to \phi_a(S) \to \phi_b(S).
\]
The specific divergence measure $\mathbb{D}$ to be used should be selected to fit the structure of the states $s \in \phi(\tilde{S})$
and may take into account the transitions $T_{\phi(\tilde{S})}$.

\subsection{Generalizing Our Model of Computation}\label{subsec:generalizing}

Before we consider specific metrics, it makes sense to first generalize our model of computation slightly.
In the real world, there will always be epistemological uncertainties in our representations and ontological
stochasticity in our implementation of any idealized computation.
We may also concern ourselves with inherently probablistic computations.
To capture this uncertainty, we will generalize our transitions in both the computational and substantial
systems to allow for probablistic transition between states.

Each transition matrix $T_{\phi(\tilde{S})}$ and $T_C$ will take the form of a transition probability matrix (TPM)
where each $T_{ij}$ is the probability that the system will transition from the $i$-th state to the $j$-th state.
Accordingly, each probability $p$ in the matrix must fall in the interval $0 \leq p \leq 1$ and must follow the normal
rules for probabilties wherein each row sums to 1.

The following is an example TPM for a system $X$ with only two states:

\[
  T_X =
\begin{bmatrix}
  P(s_0 \mid s_0) & P(s_1 \mid s_0)  \\
  P(s_0 \mid s_1) & P(s_1 \mid s_1)
\end{bmatrix}
  =
\begin{bmatrix}
  .1 & .9  \\
  .2 & .8
\end{bmatrix}
\]

This system indicates that the 0 state has a 10\% probability of remaining in the 0 state, and a 90\% probability of
transitioning to the 1 state.
The 1 state has an 80\% chance of remaining in the 1 state and a 20\% chance of transitioning to the 0 state.

\subsection{Divergence \& Optimization}\label{subsec:divergence}

There are a few choices of divergence measures that seem to present themselves fairly naturally to me that I will outline
below, but it's important to keep in mind that the best choice will depend on the nature of the representations
being utilized in $\phi({\tilde{S})}$.
I expect that there in no single correct divergence measure, but that a whole family of measures can give similar but
subtly different information about the system.

Let's take the instance where each state $s = \phi(S)$ is a binary string $\{b_0, b_1, \ldots, b_n \}$.
We can utilize earth mover's distance to define a divergence $\mathbb{D}_s$ between two states $s_a$ and $s_b$:
\[ \mathbb{D}_s\big[s_b \lVert s_a\big] = \text{EMD}(s_b,s_a). \]
Note that there are various other distance measures instead of EMD that could be used.
Now we can define a measure $\mathbb{D}_{Ts}$, which takes into account the distribution over resulting states of each
$s$:
\[ \mathbb{D}_{Ts}\big[s_b \lVert s_a\big] = \mathbb{D}_s\big[s_b \lVert s_a\big] + \frac{1}{N} \sum_i^N \mathbb{D}_s\big[T_{bi} s_i \lVert T_{ai} s_i\big]. \]

To then calculate the divergence between two representations $\phi_a(\tilde{S})$ and $\phi_b(\tilde{S})$ of equal dimension $N$, we can:
\[ \mathbb{D}_{S}\big[\phi_b(\tilde{S}) \lVert \phi_a(\tilde{S})\big] =
\sum_i^N P(s_a,s_b) \ \mathbb{D}_{Ts}\big[s_b \lVert s_a\big] \]
where $s_a = \phi_a(\tilde{S})_i$ and $s_b = \phi_b(\tilde{S})_i$, the $i$-th states of $\phi_a(\tilde{S})$ and $\phi_b(\tilde{S})$, respectively,
and $P(s_a,s_b)$ is the probability of obtaining both $s_a$ and $s_b$.

When comparing between two levels of representation $\phi_a(\tilde{S})$ and $\phi_b(\tilde{S})$, the dimensionality of their
state spaces (and therefore TPMs) may not match.
In that case, we need a model $\mathcal{M}$ that will map from the representation $\phi_a(\tilde{S})$ to $\phi_b(\tilde{S})$.
\[ \mathcal{M}\big[ \phi_a(\tilde{S}) \big] \approx \phi_b(\tilde{S}) \]
In practice, $\mathcal{M}$ may be a Hidden Markov Model (HMM) or other statistical model.

We can then indicate the divergence $\mathbb{D}_\mathcal{M}$ between levels according to the model $\mathcal{M}$ as:
\[ \mathbb{D}_\mathcal{M}\big[ \phi_a(\tilde{S}) \lVert \phi_b(\tilde{S}) \big] = \mathbb{D}_S \Big[ \mathcal{M} \big[ \phi_a(\tilde{S}) \big] \lVert \phi_b(\tilde{S}) \Big] \]

I also see great potential for introducing divergence and other metrics that take into account multiple levels of
description, for instance a divergence metric $\mathbb{D}_L$ defined over levels of description $L = \{\phi_a, \phi_b, \ldots\}$,
the simplest being:
\[ \mathbb{D}_L\big[S_2 \lVert S_1\big] = \sum_{\phi_i \in L} \mathbb{D}_{S}\big[\phi_i(S_2) \lVert \phi_i(S_1)\big]. \]
This gives a more holistic view of the difference between two systems at multiple levels of description.

We can use the divergence to select an optimal $\phi^*$ from a set of $U = \{\phi_1, \phi_2, \ldots\}$, potentially multiple
representations compatible with some computation $C$.
The following formulation selects the $\phi^*$ with the smallest divergence from a mapping $\phi_x$, potentially a lower level of description, with respect to some system $S$:
\[ \phi^* = \argmin_{\phi_i \in U} \,\mathbb{D}_S\big[\phi_i(\tilde{S}) \lVert \phi_x(\tilde{S})\big]. \]

We can also evaluate the potential $\phi_i \in U$ agaist some independently determined criteria, for example to maximize
\emph{effective information}:\footnote{See \cite{hoel2017}.}
\[ \phi^* = \argmax_{\phi_i \in U} \,\text{\textit{eff}} \big[\phi_i(\tilde{S}) \big]. \]

It seems apparent that these same techniques can apply equally as well to the computational system $C$ itself as they
do to the representation $\phi(\tilde{S})$.
For instance, we can define a divergence between a structural representation and a computational description, mediated by
a model $\mathcal{M}$:
\[ \mathbb{D}_\mathcal{M}\big[ \phi(\tilde{S}) \lVert C \big] \text{ or } \mathbb{D}_\mathcal{M}\big[ C \lVert \phi(\tilde{S}) \big]. \]

We might use a similar procedures to attribute divergences between computational systems, or to enable the selection
of an optimal $\gamma^*$ given some set $V = \{\gamma_1, \gamma_2, \ldots\}$.
With that consideration, we should be able to optimize the selection of both $\phi$ and $\gamma$.
Since these divergences are all expressed in the same units (bits) it is natural to add them together, for example:
\[ \phi^*, \gamma^* = \argmin_{\phi_i \in U, \gamma_j \in V} \mathbb{D}_S\big[\phi_i(\tilde{S}) \lVert \phi_x(\tilde{S})\big] + \mathbb{D}_\mathcal{M}\big[ \phi_i(\tilde{S}) \lVert [\gamma_j \circ \phi_i](\tilde{S}) \big]. \]

These techniques are generically compatible with other computational models, and should
be applicable in a wide array of settings.

\pagebreak

\bibliographystyle{apalike}
\bibliography{references}

\end{document}